\documentclass[12pt]{article}
\usepackage{amssymb}

\begin{document}

\bigskip\ 

\begin{center}
\textbf{LINEARIZED GRAVITY AS A GAUGE THEORY}

\bigskip\ 

\smallskip\ 

J. A. Nieto\footnote[1]{%
nieto@uas.uasnet.mx}

\smallskip\ 

\textit{Facultad de Ciencias F\'{\i}sico-Matem\'{a}ticas de la Universidad
Aut\'{o}noma}

\textit{de Sinaloa, 80010 Culiac\'{a}n Sinaloa, M\'{e}xico}

\bigskip\ 

\bigskip\ 

\textbf{Abstract}
\end{center}

We discuss linearized gravity from the point of view of a gauge theory. In
(3+1)-dimensions our analysis allows to consider linearized gravity in the
context of the MacDowell-Mansouri formalism. Our observations may be of
particular interest in the strong-weak coupling duality for linearized
gravity, in Randall-Sundrum brane world scenario and in Ashtekar formalism.

\bigskip\ 

\bigskip\ 

\bigskip\ 

\bigskip\ 

\bigskip\ 

\bigskip\ 

\bigskip\ 

\bigskip\ 


Pacs numbers: 04.20.-q, 04.60.+n, 11.15.-q, 11.10.Kk

November, 2003

\newpage

\noindent \textbf{1.- INTRODUCTION}

\smallskip\ 

Among the physical features of linearized gravity its similarity with the
source free Maxwell theory seems to be one of the most interesting. As an
example of this fact, recently it was shown [1] that the strong-weak duality
for linearized gravity makes sense as in the Maxwell theory. In this
scenario the cosmological constant, or the Planck length, plays the role of
the charge duality transformation. At this respect, it is worth mentioning
that this strong-weak duality, or S-duality, for linearized gravity seems to
have inspired other related works [2] and [3-4].

Maxwell theory is a $U(1)$ gauge theory and therefore one should expect to
be able to identify linearized gravity with some consistent gauge theory.
For instance, in a linearized version of Ashtekar formulation [5],
linearized gravity can be identified with $U(1)\times U(1)\times U(1)$
abelian gauge theory. However, this is a canonical construction and it is
more appropriate to compere it with the corresponding canonical Maxwell
theory. In this work we investigate an alternative gauge theory for
linearized gravity. Our approach has the advantage that can be applied in
the context of MacDowell-Mansouri formalism [6]. The present investigation
may be of special interest in Randall-Sundrum brane word scenario [7], in
gravitational waves formalism (see [8] and references there in) and in
quantum linearized gravity [9]. This last possibility is of particular
interest for the understanding of different aspects of quantum gravity as
has been shown by Hartle [10]. Since the quantum aspects of an abelian
theory are better understood it appears interesting to have a gauge theory
interpretation of linearized gravity.

Another source of interest in linearized gravity as a gauge theory comes
from the observation [11] that linearized gravity in four dimensions can be
understood as an irreducible representation of the de Sitter group S(4,1).
We shall show that since the cosmological constant enters in a natural way
in MacDowell-Mansouri theory, our formalism leads to the de Sitter space,
and therefore to the de Sitter group symmetry S(4,1). Moreover, since
recently, Randall-Sundrum brane world scenario [12] has motivated [13]-[16]
the study of a linearized gravity in different backgrounds and in particular
in de Sitter (or anti de Sitter) background, the present work may be
interesting in this direction.

Our work may be also useful to clarify some aspects about the relation
between the mass of the graviton and the cosmological constant which
recently has been subject of some controversy [17]-[18] in connection with
causality for the propagation of a graviton in an electromagnetic
background. This is because, once again, the cosmological constant, in the
linearized MacdDowell-Mansouri theory, arises in a natural way as we shall
show in this work.

The plan of this work is as follows: In section 2, we discuss the
traditional way to see linearized gravity as an abelian gauge theory. In
sections 3 and 4, we propose an alternative theory of linearized gravity as
gauge theory. In section 5, we apply our results to the case of linearized
MacDowell-Mansouri theory. Finally, in section 6, we make some final
comments.

\bigskip

\bigskip

\noindent \textbf{2.- LINEARIZED GRAVITY AS A GAUGE THEORY}

\smallskip\ 

Here, we closely follow reference [1]. In terms of the linearized metric%
\begin{equation}
g_{\mu \nu }=\eta _{\mu \nu }+h_{\mu \nu },  \label{1}
\end{equation}%
where $h_{\mu \nu }$ is a small deviation of the metric $g_{\mu \nu }$ from
the Minkowski metric 
\[
(\eta _{\mu \nu })=diag(-1,1,...,1), 
\]%
the first-order curvature tensor becomes%
\begin{equation}
F_{\mu \nu \alpha \beta }=-\frac{1}{2}(\partial _{\mu }\partial _{\alpha
}h_{\nu \beta }-\partial _{\mu }\partial _{\beta }h_{\nu \alpha }-\partial
_{\nu }\partial _{\alpha }h_{\mu \beta }+\partial _{\nu }\partial _{\beta
}h_{\mu \alpha }),  \label{2}
\end{equation}%
which is invariant under the transformation 
\begin{equation}
\delta h_{\mu \nu }=\partial _{\mu }\xi _{\nu }+\partial _{\nu }\xi _{\mu }.
\label{3}
\end{equation}%
Here, $\xi _{\mu }$ is an arbitrary vector field and $\partial _{\mu }\equiv 
\frac{\partial }{\partial x^{\mu }}$.

It is not difficult to see that $F_{\mu \nu \alpha \beta }$ satisfies the
following relations: 
\begin{equation}
\begin{array}{c}
F_{\mu \nu \alpha \beta }=-F_{\mu \nu \beta \alpha }=-F_{\nu \mu \alpha
\beta }=F_{\alpha \beta \mu \nu }, \\ 
\\ 
F_{\mu \nu \alpha \beta }+F_{\mu \beta \nu \alpha }+F_{\mu \alpha \beta \nu
}=0, \\ 
\\ 
\partial _{\lambda }F_{\mu \nu \alpha \beta }+\partial _{\mu }F_{\nu \lambda
\alpha \beta }+\partial _{\nu }F_{\lambda \mu \alpha \beta }=0.%
\end{array}
\label{4}
\end{equation}%
It is interesting to observe that, in 3+1 dimensions, the dual $^{\ast
}F_{\mu \nu \alpha \beta }\equiv \frac{1}{2}\epsilon _{\mu \nu \tau \sigma
}F_{\alpha \beta }^{\tau \sigma },$ where $\epsilon _{\mu \nu \alpha \beta }$
is a completely antisymmetric Levi-Civita tensor with $\epsilon _{0123}=-1$
and the indices are raised and lowered by means of $\eta ^{\alpha \beta }$
and $\eta _{\alpha \beta },$ does not satisfy the relations (4) unless $%
F_{\mu \alpha \nu }^{\alpha }$ satisfies the vacuum Einstein equations $%
F_{\mu \nu }=F_{\mu \alpha \nu }^{\alpha }=0$.

Let us introduce the `gauge' field 
\begin{equation}
A_{\mu \alpha \beta }=\frac{1}{2}(\partial _{\beta }h_{\mu \alpha }-\partial
_{\alpha }h_{\mu \beta }).  \label{5}
\end{equation}%
Observe that $A_{\mu \alpha \beta }=-A_{\mu \beta \alpha }.$ Using (3), we
find that $A_{\mu \alpha \beta }$ transforms as 
\begin{equation}
\delta A_{\mu \alpha \beta }=\partial _{\mu }\lambda _{\alpha \beta },
\label{6}
\end{equation}%
where $\lambda _{\alpha \beta }=\frac{1}{2}(\partial _{\beta }\xi _{\alpha
}-\partial _{\alpha }\xi _{\beta })=-\lambda _{\beta \alpha }$.

The expression (5) allows to write the curvature tensor $F_{\mu \nu
}^{\alpha \beta }$ as 
\begin{equation}
F_{\mu \nu }^{\alpha \beta }=\partial _{\mu }A_{\nu }^{\alpha \beta
}-\partial _{\nu }A_{\mu }^{\alpha \beta }.  \label{7}
\end{equation}%
Thus, we have shown that the tensor $F_{\mu \nu }^{\alpha \beta }$ may be
written in the typical form of an abelian Maxwell field strength $F_{\mu \nu
}^{a}=\partial _{\mu }A_{\nu }^{a}-\partial _{\nu }A_{\mu }^{a},$ where the
index $a$ runs over some abelian group such as $U(1)\times U(1)\times
...\times U(1)$. Note that $F_{\mu \nu }^{\alpha \beta }$ is invariant under
the transformation $\delta A_{\mu }^{\alpha \beta }=\partial _{\mu }\lambda
^{\alpha \beta }$ which has exactly the same form as the transformation of
abelian gauge fields $\delta A_{\mu }^{a}=\partial _{\mu }\lambda ^{a},$
where $\lambda ^{a}$ is an arbitrary function of the coordinates $x^{\mu }$.

\smallskip\ 

\noindent \textbf{3.- ALTERNATIVE DESCRIPTION OF LINEARIZED GRAVITY AS A
GAUGE THEORY}

\smallskip\ 

The motivation for the present work arose from the observation that although 
$F_{\mu \nu }^{\alpha \beta }$ has the typical form of an abelian field
strength, $A_{\mu \alpha \beta }$ has also such a form. In fact, some
authors (see [17] and references therein) take the combination $A_{\mu
\alpha \beta }A^{\mu \alpha \beta }$ as part of the kinetic term in the
Fierz-Pauli action [19]. In what follows we shall show that in fact $F_{\mu
\nu }^{\alpha \beta }$ and $A_{\mu \alpha \beta }$ can be associated to the
curvature of a gauge field, but while $F_{\mu \nu }^{\alpha \beta }$ can be
identified with the genuine reduced curvature, $A_{\mu \alpha \beta }$ is
related to the torsion. We shall also clarify the group aspects associated
to $F_{\mu \nu }^{\alpha \beta }$ and $A_{\mu \alpha \beta }$.

Consider a $SO(n,1)$ gravitational gauge field $\omega _{\mu }^{AB}=-$ $%
\omega _{\mu }^{BA}$, where the index $\mu $ runs from $0$ to $n-1.$ Assume
that such a gauge field is broken into a $SO(n-1,1)$ gauge field connection $%
\omega _{\mu }^{ab}$ and the $\omega _{\mu }^{na}=e_{\mu }^{a}$ vierbein
field, with $n$ as a fixed index With these notation we find that the $%
SO(n,1)$ curvature

\begin{equation}
\mathcal{R}_{\mu \nu }^{AB}=\partial _{\mu }\omega _{\nu }^{AB}-\partial
_{\nu }\omega _{\mu }^{AB}+\omega _{\mu }^{AC}\omega _{\nu C}^{\quad
B}-\omega _{\nu }^{AC}\omega _{\mu C}^{\quad B}  \label{8}
\end{equation}%
leads to 
\begin{equation}
\mathcal{R}_{\mu \nu }^{ab}=R_{\mu \nu }^{ab}-\Sigma _{\mu \nu }^{ab}
\label{9}
\end{equation}%
and 
\begin{equation}
\mathcal{R}_{\mu \nu }^{na}=\partial _{\mu }e_{\nu }^{a}-\partial _{\nu
}e_{\mu }^{a}+\omega _{\mu }^{ac}e_{\nu c}-\omega _{\nu }^{ac}e_{\mu c},
\label{10}
\end{equation}%
where 
\begin{equation}
R_{\mu \nu }^{ab}=\partial _{\mu }\omega _{\nu }^{ab}-\partial _{\nu }\omega
_{\mu }^{ab}+\omega _{\mu }^{ac}\omega _{\nu c}^{\quad b}-\omega _{\nu
}^{ac}\omega _{\mu c}^{\quad b}  \label{11}
\end{equation}%
is the $SO(n-1,1)$ curvature and 
\begin{equation}
\Sigma _{\mu \nu }^{ab}=e_{\mu }^{a}e_{\nu }^{b}-e_{\nu }^{a}e_{\mu }^{b}.
\label{12}
\end{equation}%
It turns out that $T_{\mu \nu }^{a}\equiv \mathcal{R}_{\mu \nu }^{na}$ can
be identified with the torsion.

Let us now assume a vanishing torsion; $T_{\mu \nu }^{a}=0.$ From (10) it
follows that

\begin{equation}
\partial _{\mu }e_{\nu }^{a}-\Gamma _{\mu \nu }^{\alpha }e_{\alpha
}^{a}+\omega _{\mu }^{ac}e_{\nu c}=0,  \label{13}
\end{equation}%
where $\Gamma _{\mu \nu }^{\alpha }=\Gamma _{\nu \mu }^{\alpha }$. Using
(13) we find that $\omega _{\mu \nu \alpha }=e_{\nu a}e_{\alpha b}\omega
_{\mu }^{ab}$ becomes

\begin{equation}
\omega _{\mu \nu \alpha }=\frac{1}{2}(e_{\alpha a}\Lambda _{\mu \nu
}^{a}+e_{\mu a}\Lambda _{\alpha \nu }^{a}+e_{\nu a}\Lambda _{\alpha \mu
}^{a}),  \label{14}
\end{equation}%
where

\begin{equation}
\Lambda _{\mu \nu }^{a}=\partial _{\mu }e_{\nu }^{a}-\partial _{\nu }e_{\mu
}^{a}.  \label{15}
\end{equation}

The viernbein field $e_{\mu }^{a}$ and the metric $g_{\mu \nu }$ can be
related through the formula

\begin{equation}
g_{\mu \nu }=e_{\mu }^{a}e_{\mu }^{b}\eta _{ab},  \label{16}
\end{equation}%
where $\eta _{ab}$ is an internal Minkowski metric.

So far we have not considered any approximation. In the linear case, we can
write $e_{\mu }^{a}=b_{\mu }^{a}+l_{\mu }^{a}.$ Thus, considering that $\eta
_{\mu \nu }=b_{\mu }^{a}b_{\nu }^{b}\eta _{ab}$, at first order the
expression (16) becomes

\begin{equation}
g_{\mu \nu }=\eta _{\mu \nu }+b_{\mu }^{a}l_{\nu }^{b}\eta _{ab}+b_{\nu
}^{a}l_{\mu }^{b}\eta _{ab}.  \label{17}
\end{equation}%
Therefore, by comparing this expression with (1) we find

\begin{equation}
h_{\mu \nu }=l_{\mu \nu }+l_{\nu \mu }.  \label{18}
\end{equation}%
Here, we used the definition $l_{\mu \nu }\equiv b_{\nu }^{a}l_{\mu
}^{b}\eta _{ab}.$ To first order the formula (14) becomes

\begin{equation}
\omega _{\mu \nu \alpha }=\frac{1}{2}(b_{\alpha a}(\partial _{\mu }l_{\nu
}^{a}-\partial _{\nu }l_{\mu }^{a})+b_{\mu a}(\partial _{\alpha }l_{\nu
}^{a}-\partial _{\nu }l_{\alpha }^{a})+b_{\nu a}(\partial _{\alpha }l_{\mu
}^{a}-\partial _{\mu }l_{\alpha }^{a})).  \label{19}
\end{equation}%
Considering that $\partial _{\mu }b_{\alpha a}=0$ and that $l_{\mu \nu
}=b_{\nu a}l_{\mu }^{a}$ we find that this expression can also be written as

\begin{equation}
\omega _{\mu \nu \alpha }=\frac{1}{2}((\partial _{\mu }l_{\nu \alpha
}-\partial _{\nu }l_{\mu \alpha })+(\partial _{\alpha }l_{\nu \mu }-\partial
_{\nu }l_{\alpha \mu })+(\partial _{\alpha }l_{\mu \nu }-\partial _{\mu
}l_{\alpha \nu })).  \label{20}
\end{equation}

Now using (5) and (18) we discover that (20) can be written as

\begin{equation}
\omega _{\mu \nu \alpha }=A_{\mu \nu \alpha }+\frac{1}{2}\partial _{\mu
}f_{\nu \alpha },  \label{21}
\end{equation}%
where

\begin{equation}
f_{\nu \alpha }=(l_{\nu \alpha }-l_{\alpha \nu }).  \label{22}
\end{equation}%
Therefore, we have shown that up to a gauge transformation $\omega _{\mu \nu
\alpha }$ and $A_{\mu \nu \alpha }$ describe the same physics. From (7), we
also find that to first order

\begin{equation}
R_{\mu \nu }^{ab}=b_{\alpha }^{a}b_{\beta }^{b}F_{\mu \nu }^{\alpha \beta }.
\label{23}
\end{equation}%
With these results at hand we observe that the reason that $A_{\mu \nu
\alpha }$ has the form given in (5) is because the torsion vanishes. Now
that we have the links $\omega _{\mu \nu \alpha }\leftrightarrow A_{\mu \nu
\alpha }$ and $R_{\mu \nu }^{ab}\leftrightarrow F_{\mu \nu }^{\alpha \beta }$
we may proceed to analyze the group structure underling such links.

\smallskip\ 

\noindent \textbf{4.- GROUP STRUCTURE IN LINEARIZED GRAVITY}

\smallskip\ 

In order to capture the group information in the development of the previous
section we introduce the generator $J_{AB}=-J_{BA}$ associated to the group $%
SO(n,1)$ and we write $\omega _{\mu }$ and $\mathcal{R}_{\mu \nu }$ as

\begin{equation}
\omega _{\mu }=\frac{1}{2}\omega _{\mu }^{AB}J_{AB}  \label{24}
\end{equation}%
and

\begin{equation}
\mathcal{R}_{\mu \nu }=\frac{1}{2}\mathcal{R}_{\mu \nu }^{AB}J_{AB},
\label{25}
\end{equation}%
respectively.

It turns out that $\mathcal{R}_{\mu \nu }$ can be written in terms of $%
\omega _{\mu }$ in the form

\begin{equation}
\mathcal{R}_{\mu \nu }=\partial _{\mu }\omega _{\nu }-\partial _{\nu }\omega
_{\mu }+\frac{1}{4}[\omega _{\mu },\omega _{\nu }],  \label{26}
\end{equation}%
where $[\omega _{\mu },\omega _{\nu }]$ is determined by the Lie algebra

\begin{equation}
\lbrack J_{AB},J_{CD}]=\frac{1}{2}C_{ABCD}^{EF}J_{EF}.  \label{27}
\end{equation}%
Here, $C_{ABCD}^{EF}$ are the structure constants associated to $SO(n,1)$.
This algebra can be written as

\begin{equation}
\lbrack J_{ab},J_{cd}]=\frac{1}{2}C_{abcd}^{ef}J_{ef},  \label{28}
\end{equation}%
\begin{equation}
\lbrack J_{na},J_{cd}]=C_{nacd}^{nf}J_{nf}=\delta _{ac}J_{nd}-\delta
_{ad}J_{nc},  \label{29}
\end{equation}%
\begin{equation}
\lbrack J_{na},J_{nb}]=\frac{1}{2}C_{nanb}^{ef}J_{ef}=J_{ab}.  \label{30}
\end{equation}

Using (28)-(30) we find that (26) becomes

\begin{equation}
\begin{array}{c}
\mathcal{R}_{\mu \nu }=\frac{1}{2}(\partial _{\mu }\omega _{\nu
}^{ab}-\partial _{\nu }\omega _{\mu }^{ab})J_{ab}+(\partial _{\mu }\omega
_{\nu }^{nb}-\partial _{\nu }\omega _{\mu }^{nb})J_{nb} \\ 
\\ 
+\frac{1}{4}[J_{ab},J_{cd}]\omega _{\mu }^{ab}\omega _{\nu }^{cd}+\frac{1}{2}%
[J_{na},J_{cd}](\omega _{\mu }^{na}\omega _{\nu }^{cd}-\omega _{\mu
}^{cd}\omega _{\nu }^{na}) \\ 
\\ 
+[J_{na},J_{nb}]\omega _{\mu }^{na}\omega _{\nu }^{nb}.%
\end{array}
\label{31}
\end{equation}%
In order to study the linearized approximation of (31) let us write $\omega
_{\mu }^{ab}\rightarrow \frac{1}{\lambda }\omega _{\mu }^{ab}$ and $\omega
_{\mu }^{na}\rightarrow b_{\mu }^{a}+\frac{1}{\lambda }l_{\mu }^{a}$ where $%
\lambda $ is an auxiliary parameter measuring the order of approximation.
The expression (31) becomes

\begin{equation}
\begin{array}{c}
\mathcal{R}_{\mu \nu }=\frac{1}{2\lambda }(\partial _{\mu }\omega _{\nu
}^{ab}-\partial _{\nu }\omega _{\mu }^{ab})J_{ab}+\frac{1}{\lambda }%
(\partial _{\mu }l_{\nu }^{a}-\partial _{\nu }l_{\mu }^{a})J_{na} \\ 
\\ 
+\frac{1}{4\lambda ^{2}}[J_{ab},J_{cd}]\omega _{\mu }^{ab}\omega _{\nu
}^{cd}+\frac{1}{2\lambda }[J_{na},J_{cd}](b_{\mu }^{a}\omega _{\nu
}^{cd}-\omega _{\mu }^{cd}b_{\nu }^{a})+\frac{1}{2\lambda ^{2}}%
[J_{na},J_{cd}](l_{\mu }^{a}\omega _{\nu }^{cd}-\omega _{\mu }^{cd}l_{\nu
}^{a}) \\ 
\\ 
+[J_{na},J_{nb}]b_{\mu }^{a}b_{\nu }^{b}+\frac{1}{\lambda }%
[J_{na},J_{nb}](b_{\mu }^{a}l_{\nu }^{b}-l_{\mu }^{b}b_{\nu }^{a})+\frac{1}{%
\lambda ^{2}}[J_{na},J_{nb}]l_{\mu }^{a}l_{\nu }^{b}.%
\end{array}
\label{32}
\end{equation}%
From this expression it is evident that when $\lambda \rightarrow \infty $
the term $\frac{1}{4\lambda ^{2}}[J_{ab},J_{cd}]\omega _{\mu }^{ab}\omega
_{\nu }^{cd}$ may be interpreted as a Wigner contraction leading to $%
[J_{ab},J_{cd}]=0$ in the limit $\lambda \rightarrow \infty $ and therefore
the group structure with respect to this bracket in the limit $\lambda
\rightarrow \infty $ corresponds to an abelian theory. However, we observe
that since there are both factors $\frac{1}{\lambda }$ and $\frac{1}{\lambda
^{2}}$ in connection with the brackets $[J_{na},J_{cd}]$ and $%
[J_{na},J_{nb}] $, it is not possible to apply a Wigner procedure to these
two brackets and therefore the theory do not admit the interpretation of an
abelian theory with respect these two brackets. From these observations it
follows that in the limit $\lambda \rightarrow \infty $ the algebra
(28)-(30) can be taken as

\begin{equation}
\lbrack J_{ab},J_{cd}]=0,  \label{33}
\end{equation}%
\begin{equation}
\lbrack J_{na},J_{cd}]=C_{nacd}^{nf}J_{nf}=\delta _{ac}J_{nd}-\delta
_{ad}J_{nc},  \label{34}
\end{equation}%
\begin{equation}
\lbrack J_{na},J_{nb}]=\frac{1}{2}C_{nanb}^{ef}J_{ef}=J_{ab}.  \label{35}
\end{equation}%
Thus, we have shown that it is consistent to substitute from the beginning
the algebra (33)-(35) into (32) and then to proceed to eliminate terms of
order $\frac{1}{\lambda ^{2}}.$ In a essence what we have proved is that in
the limit $\lambda \rightarrow \infty $ it makes sense to take the subgroup $%
SO(n-1,1)$ of $SO(n,1)$ as a `collapsed' abelian group.

\smallskip\ 

\noindent \textbf{5.- THE ACTION STRUCTURE FOR A GAUGE THEORY OF LINEARIZED
GRAVITY IN FOUR DIMENSIONS}

\smallskip\ 

So far we have analyzed linearized gravity at the level of gauge field and
its correspondent curvature. In order to complete our analysis we need to
introduce an action. In this scenario the Lovelock theory [20] seems to be
the most indicated for our goal. Just to avoid unnecessary complications in
the general case, here, we focus on the Lovelock theory in four dimensions
which has become known as a MacDowell-Mansouri theory [6].

In four dimensions the gauge group $SO(n,1)$ becomes the de Sitter group $%
SO(4,1)$. In this case, the gravitational gauge field $\omega _{\mu }^{AB}$
is broken into the $SO(3,1)$ connection $\omega _{\mu }^{ab}$ and the $%
\omega _{\mu }^{4a}=e_{\mu }^{a}$ tetrad field. Thus, in the weak field
approximation, the $SO(4,1)$ de Sitter curvature (8) leads to 
\begin{equation}
\mathcal{R}_{\mu \nu }^{ab}=R_{\mu \nu }^{ab}-\Sigma _{\mu \nu }^{ab}
\label{36}
\end{equation}%
and 
\begin{equation}
\mathcal{R}_{\mu \nu }^{4a}=\partial _{\mu }l_{\nu }^{a}-\partial _{\nu
}l_{\mu }^{a}+\omega _{\mu }^{ac}b_{\nu c}-\omega _{\nu }^{ac}b_{\mu c},
\label{37}
\end{equation}%
where 
\begin{equation}
R_{\mu \nu }^{ab}=\partial _{\mu }\omega _{\nu }^{ab}-\partial _{\nu }\omega
_{\mu }^{ab}  \label{38}
\end{equation}%
is the linearized curvature and

\begin{equation}
\Sigma _{\mu \nu }^{ab}=b_{\mu }^{a}b_{\nu }^{b}-b_{\nu }^{a}b_{\mu
}^{b}+b_{\mu }^{a}l_{\nu }^{b}-b_{\nu }^{a}l_{\mu }^{b}+l_{\mu }^{a}b_{\nu
}^{b}-l_{\nu }^{a}b_{\mu }^{b}.  \label{39}
\end{equation}%
Here, in the computation of $\Sigma _{\mu \nu }^{ab}$, we have used the
relation $e_{\mu }^{a}=b_{\mu }^{a}+l_{\mu }^{a}$ and the approximation $%
l_{\mu }^{a}l_{\nu }^{b}=l_{\nu }^{a}l_{\mu }^{b}=0$. From (39) we find that 
$\Sigma _{\mu \nu }^{\alpha \beta }=b_{a}^{\alpha }b_{b}^{\beta }\Sigma
_{\mu \nu }^{ab}$ is given by

\begin{equation}
\Sigma _{\mu \nu }^{\alpha \beta }=\delta _{\mu \nu }^{\alpha \beta }+\Omega
_{\mu \nu }^{\alpha \beta },  \label{40}
\end{equation}%
where

\begin{equation}
\delta _{\mu \nu }^{\alpha \beta }=\delta _{\mu }^{\alpha }\delta _{\nu
}^{\beta }-\delta _{\mu }^{\beta }\delta _{\nu }^{\alpha }  \label{41}
\end{equation}%
and

\begin{equation}
\Omega _{\mu \nu }^{\alpha \beta }=\delta _{\mu }^{\alpha }l_{\nu }^{\beta
}-\delta _{\nu }^{\alpha }l_{\mu }^{\beta }+l_{\mu }^{\alpha }\delta _{\nu
}^{\beta }-l_{\nu }^{\alpha }\delta _{\mu }^{\beta }.  \label{42}
\end{equation}

The MacDowell-Mansouri's action is 
\begin{equation}
S=\frac{1}{4}\int d^{4}x{}\varepsilon ^{\mu \nu \alpha \beta }{}\mathcal{R}%
_{\mu \nu }^{ab}{}\mathcal{R}_{\alpha \beta }^{cd}{}\epsilon _{abcd},
\label{43}
\end{equation}%
where $\varepsilon ^{\mu \nu \alpha \beta }$ is the completely antisymmetric
tensor associated to the space-time, with $\varepsilon ^{0123}=1$ and $%
\varepsilon _{0123}=1$, while $\epsilon _{abcd}$ is also the completely
antisymmetric tensor but now associated to the internal group $S(3,1)$, with 
$\epsilon _{0123}=-1.$ We assume that the internal metric is given by $(\eta
_{ab})=(-1,1,1,1).$ Therefore, we have $\epsilon ^{0123}=1.$ It is well
known that, in the general case, the action (43) leads to three terms; the
Hilbert Einstein action, the cosmological constant term and the Euler
topological invariant (or Gauss-Bonnet term). It is worth mentioning that
the action (43) may also be obtained considering Lovelock theory [20] in
four dimensions.

In the linearized case the action (43) can be written as

\begin{equation}
S=\frac{1}{4}\int d^{4}x{}\epsilon ^{\mu \nu \alpha \beta }{}\mathcal{R}%
_{\mu \nu }^{\alpha \beta }{}\mathcal{R}_{\alpha \beta }^{\rho \sigma
}{}\epsilon _{\alpha \beta \rho \sigma },  \label{44}
\end{equation}%
with $\mathcal{R}_{\mu \nu }^{\alpha \beta }=b_{a}^{\alpha }b_{b}^{\beta }%
\mathcal{R}_{\mu \nu }^{ab}$. Using (36) and (40) we find

\begin{equation}
\begin{array}{c}
S=\frac{1}{4}\int d^{4}x{}\epsilon ^{\mu \nu \alpha \beta }R_{\mu \nu
}^{\gamma \delta }R_{\alpha \beta }^{\rho \sigma }\epsilon _{\gamma \delta
\rho \sigma }-\frac{1}{2}\int d^{4}x{}\epsilon ^{\mu \nu \alpha \beta
}\delta _{\mu \nu }^{\gamma \delta }R_{\alpha \beta }^{\rho \sigma }\epsilon
_{\gamma \delta \rho \sigma } \\ 
\\ 
-\frac{1}{2}\int d^{4}x{}\epsilon ^{\mu \nu \alpha \beta }\Omega _{\mu \nu
}^{\gamma \delta }R_{\alpha \beta }^{\rho \sigma }\epsilon _{\gamma \delta
\rho \sigma }+\frac{1}{4}\int d^{4}x{}\epsilon ^{\mu \nu \alpha \beta
}\delta _{\mu \nu }^{\gamma \delta }\delta _{\alpha \beta }^{\rho \sigma
}\epsilon _{\gamma \delta \rho \sigma } \\ 
\\ 
+\frac{1}{2}\int d^{4}x{}\epsilon ^{\mu \nu \alpha \beta }\Omega _{\mu \nu
}^{\gamma \delta }\delta _{\alpha \beta }^{\rho \sigma }\epsilon _{\gamma
\delta \rho \sigma }+\frac{1}{4}\int d^{4}x{}\epsilon ^{\mu \nu \alpha \beta
}\Omega _{\mu \nu }^{\gamma \delta }\Omega _{\alpha \beta }^{\rho \sigma
}\epsilon _{\gamma \delta \rho \sigma }.%
\end{array}
\label{45}
\end{equation}%
From (45) and (38) we observe that the second terms is total derivatives.
The fourth term is just a constant and therefore can be drooped from the
action. The fifth term is proportional to $tr(l_{\beta }^{\alpha })=\frac{1}{%
2}tr(h_{\beta }^{\alpha })$ and can be dropped in a gauge fixing $%
tr(h_{\beta }^{\alpha })=0$. The first term is also a total derivative but
it is useful to keep it for S-duality considerations [1]. Therefore,
considering the dynamic and topological important terms the action (45) is
reduced to

\begin{equation}
\begin{array}{c}
S=\frac{1}{4}\int d^{4}x{}\epsilon ^{\mu \nu \alpha \beta }R_{\mu \nu
}^{\gamma \delta }R_{\alpha \beta }^{\rho \sigma }\epsilon _{\gamma \delta
\rho \sigma }-\frac{1}{2}\int d^{4}x{}\epsilon ^{\mu \nu \alpha \beta
}\Omega _{\mu \nu }^{\gamma \delta }R_{\alpha \beta }^{\rho \sigma }\epsilon
_{\gamma \delta \rho \sigma } \\ 
\\ 
+\frac{1}{4}\int d^{4}x{}\varepsilon ^{\mu \nu \alpha \beta }\Omega _{\mu
\nu }^{\gamma \delta }\Omega _{\alpha \beta }^{\rho \sigma }\epsilon
_{\gamma \delta \rho \sigma }.%
\end{array}
\label{46}
\end{equation}%
This action was the starting point in reference [1] in connection with the
S-duality for linearized gravity. Using (41) and (42) we find 
\begin{equation}
\begin{array}{c}
S=\frac{1}{4}\int d^{4}x{}\epsilon ^{\mu \nu \alpha \beta }R_{\mu \nu
}^{\gamma \delta }R_{\alpha \beta }^{\rho \sigma }\epsilon _{\gamma \delta
\rho \sigma }-2\int d^{4}x{}\epsilon ^{\mu \nu \alpha \beta }{}\delta _{\mu
}^{\tau }{}l_{\nu }^{\lambda }R_{\alpha \beta }^{\sigma \rho }{}\epsilon
_{\tau \lambda \sigma \rho } \\ 
\\ 
+4\int d^{4}x{}\epsilon ^{\mu \nu \alpha \beta }{}\delta _{\mu }^{\tau
}l_{\nu }^{\lambda }{}\delta _{\alpha }^{\sigma }l_{\beta }^{\rho
}{}\epsilon _{\tau \lambda \sigma \rho }.%
\end{array}
\label{47}
\end{equation}%
Since $\epsilon ^{\mu \nu \alpha \beta }{}\delta _{\mu }^{\tau }{}\epsilon
_{\tau \lambda \sigma \rho }=-\delta _{\lambda \sigma \rho }^{\nu \alpha
\beta }$ and $\epsilon ^{\mu \nu \alpha \beta }{}\delta _{\mu }^{\tau
}{}\delta _{\alpha }^{\sigma }{}\epsilon _{\tau \lambda \sigma \rho
}=-2{}\delta _{\lambda \rho }^{\nu \beta }$, where in general $\delta _{\tau
\lambda \sigma \rho }^{\mu \nu \alpha \beta }$ is a generalized delta, we
discover that (47) can be written as 
\begin{equation}
\begin{array}{c}
S=\frac{1}{4}\int d^{4}x{}\epsilon ^{\mu \nu \alpha \beta }{}R_{\mu \nu
}^{\tau \lambda }{}R_{\alpha \beta }^{\sigma \rho }{}\epsilon _{\tau \lambda
\sigma \rho }-4\int d^{4}x{}h^{\mu \nu }(R_{\mu \nu }-\frac{1}{2}\eta _{\mu
\nu }R) \\ 
\\ 
-2\int d^{4}x{}(h^{2}-h^{\mu \nu }h_{\mu \nu })-2\int d^{4}x{}f^{\mu \nu
}f_{\mu \nu }).%
\end{array}
\label{48}
\end{equation}%
Here, we used the following definitions: $R_{\mu \nu }\equiv \eta ^{\alpha
\beta }R_{\mu \alpha \nu \beta },R\equiv \eta ^{\mu \nu }\eta ^{\alpha \beta
}R_{\mu \alpha \nu \beta }$ and $h\equiv \eta ^{\mu \nu }h_{\mu \nu }=2\eta
^{\mu \nu }l_{\mu \nu }.$ Further, we considered that $R_{\mu \nu }=R_{\nu
\mu }$ and $h^{\mu \nu }R_{\mu \nu }=2l^{\mu \nu }R_{\mu \nu }.$ We
recognize the second term and the third term in (48) as the Einstein action
for linearized gravity with cosmological constant, while the first term is a
total derivative. The last term in (48) becomes a total derivative under the
transformation $\delta h_{\mu \nu }=\partial _{\mu }\xi _{\nu }+\partial
_{\nu }\xi _{\mu },$ given in (3), since in this case according to (6) and
(21) one may have $f_{\mu \nu }=(\partial _{\mu }\xi _{\nu }-\partial _{\nu
}\xi _{\mu })$.

It is worth mentioning that in order to study S-duality for linearized
gravity the action (44) was generalized in the form [1]

\begin{equation}
\mathcal{S}=\frac{1}{4}(^{+}\tau )\int d^{4}x{}\epsilon ^{\mu \nu \alpha
\beta }{}^{+}\mathcal{R}_{\mu \nu }^{\tau \lambda }{}^{+}\mathcal{R}_{\alpha
\beta }^{\sigma \rho }{}\epsilon _{\tau \lambda \sigma \rho }-\frac{1}{4}%
(^{-}\tau )\int d^{4}x{}\epsilon ^{\mu \nu \alpha \beta }{}^{-}\mathcal{R}%
_{\mu \nu }^{\tau \lambda }{}^{-}\mathcal{R}_{\alpha \beta }^{\sigma \rho
}{}\epsilon _{\tau \lambda \sigma \rho },  \label{49}
\end{equation}%
where $^{+}\tau $ and $^{-}\tau $ are two different constant parameters and $%
^{\pm }\mathcal{F}_{\mu \nu }^{\alpha \beta }$ is given by 
\begin{equation}
^{\pm }\mathcal{F}_{\mu \nu }^{\alpha \beta }=(\frac{1}{2})^{\pm }M_{\tau
\lambda }^{\alpha \beta }\mathcal{F}_{\mu \nu }^{\tau \lambda },  \label{50}
\end{equation}%
where 
\begin{equation}
^{\pm }M_{\tau \lambda }^{\alpha \beta }=\frac{1}{2}(\delta _{\tau \lambda
}^{\alpha \beta }\mp i\epsilon _{\quad \tau \lambda }^{\alpha \beta }).
\label{51}
\end{equation}%
It turns out that $^{+}\mathcal{F}_{\mu \nu }^{\alpha \beta }$ is self-dual,
while $^{-}\mathcal{F}_{\mu \nu }^{\alpha \beta }$ is anti self-dual
curvature. Therefore, the action (49) describes self-dual and anti-self-dual
linearized gravity.

\smallskip\ 

\noindent \textbf{6.-FINAL COMMENTS}

\smallskip\ 

In this article, we have investigated different aspects of linearized
gravity as a gauge theory. We showed that linearized gravity can be
understood as an abelian gauge theory only in connection with the subgroup $%
S(n-1,1)$ but not in connection with the full group $S(n,1)$. We apply our
observations to the case of MacDowell-Mansouri theory in four dimensions
showing that our proposed method leads to the Fierz-Pauli action in the weak
field limit. Furthermore, we argued that our results can be directly used in
the case of S-duality for linearized gravity.

In fact, in reference [1] it was discovered that the S-duality for
linearized gravity leads to the cosmological constant duality $\Lambda
\rightarrow \frac{1}{\Lambda },$ in a similar form as an abelian gauge
theory leads to the duality of the gauge coupling constant $g^{2}\rightarrow 
\frac{1}{g^{2}}.$ The results of the present work seems to confirm this
analogy since we showed that in a strictly sense linearized gravity can be
in fact understood as an Abelian gauge theory. Moreover, it is evident from
reference [1] that instead of the duality $\Lambda \rightarrow \frac{1}{%
\Lambda }$ it is possible to consider the duality symmetry $l_{p}\rightarrow 
\frac{1}{l_{p}}$ (or the more general duality transformation $l_{p}\Lambda
\rightarrow \frac{1}{\Lambda l_{p}})$. It is worth mentioning that recently,
there have been much interest in the strong coupling limit of linearized
gravity [4] via the duality $l_{p}\rightarrow \frac{1}{l_{p}}$. It may be
interesting for further research to see if our present study is useful in
this direction.

Other possible application of the results of the present work is in
connection with the non-minimal coupling for spin 3/2 field studied in
reference [21]. In this reference the authors find the non-minimal coupling
for spin 3/2 by applying the method of `square root' to the constrains of
spin 2 gauge field [22]. These constrains are obtained considering the
traditional method of linearized gravity presented in section 2. Now, that
we have found the route to see linearized gravity as a gauge theory it seems
interesting to revisited the coupling for spin 3/2 field. This possibility
may shed some light on quantum linearized supergravity and since
supergravity is an essential part of M-theory [23-25] one should expect
eventually to gain some better understanding of such a theory. This seems a
viable route to follow because it has been proved [26] that massive spin 2
coupled to gravity is deeply connected to string theory.

Finally it has been studied [27] the role of little group in linearized
gravity obtaining the result that the translational subgroup of the Wigner's
little group acts as a generator of linearized gravity only when the space
time has dimension four. It may be interesting for further research to
investigate this intriguing result from the point of view of the present
work.

\bigskip\ 

\noindent \textbf{Acknowledgments}

\noindent I would like to thank to M. C. Mar\'{\i}n for helpful comments.

\end{document}